\documentclass[aps,twocolumn]{revtex4}
\makeatletter

\newcommand{\Rmnum}[1]{\expandafter\@slowromancap\romannumeral #1@}
\makeatother
\usepackage{graphicx}
\usepackage{dcolumn}
\usepackage{bm}
\usepackage{CJK}
\usepackage{amsfonts}
\usepackage{pifont}
\usepackage{psfrag}
\usepackage{wrapfig}
\usepackage{subfigure}
\usepackage{makeidx}
\usepackage{multirow}
\usepackage{epsf}
\usepackage[colorlinks,linkcolor=red,anchorcolor=red,citecolor=blue,urlcolor=blue]{hyperref}
\usepackage{amsmath}
\usepackage{cases}
\usepackage{amssymb}
\usepackage{soul}
\usepackage[version=3]{mhchem}

\usepackage{color}
\setulcolor{red}

\begin{document}
\begin{CJK}{GBK}{song}
\title{The topological phase of bright solitons}

\author{Yu-Han Wu$^{1,2,3}$}
\author{Li-Chen Zhao$^{1,2,3,5}$}
\author{Chong Liu$^{1,2,3,6}$}
\author{Zhan-Ying Yang$^{1,2,3}$}
\author{Wen-Li Yang$^{1,2,3,4}$}
\address{$^1$School of Physics, Northwest University, Xi'an 710127, China}
\address{$^2$Shaanxi Key Laboratory for Theoretical Physics Frontiers, Xi'an 710127, China}
\address{$^3$NSFC-SPTP Peng Huanwu Center for Fundamental Theory, Xi'an 710127, China}
\address{$^4$Institute of Modern Physics, Northwest University, Xi'an 710127}
\address{$^5$e-mail: zhaolichen3@nwu.edu.cn}
\address{$^6$e-mail: chongliu@nwu.edu.cn}

\begin{abstract}
We study the topological phase of bright soliton with arbitrary velocity under the self-steepening effect. Such topological phase can be described by the topological vector potential and effective magnetic field. We find that the point-like magnetic fields corresponds to the density peak of such bright solitons, where each elementary magnetic flux is $\pi$. Remarkably, we show that two bright solitons can generate an additional topological field due to the phase jump between them. Our research provided the possibility to use bright solitons to explore topological properties.
\end{abstract}

\maketitle

Topological solitons, are stable, particle-like objects with finite mass and a smooth structure, which has been studied in many fields of physics including the nuclear physics \cite{nuclear1,nuclear2}, superconductors \cite{superC1,superC2}, quantum field theory and string theory \cite{quantum,string}. The remarkable properties of these topological solitons can be characterized by the Aharonov-Bohm effect \cite{ABef}, Berry phase \cite{Berry}, and so on \cite{quanHall}. So far, most of the studies are limited to the N-dimensional $(N>1)$ systems admitting localized topological solitons. Examples are instantons in four dimensions \cite{superC2,string,instanton1,instanton2}, knots and skyrmions in three dimensions \cite{knot,nuclear1,skyr,instanton2}, and vortices in two dimension \cite{vortex1,vortex2}.

As the one-dimensional (1-D) counterpart of the two-dimensional (2-D) vortices, the dark solitons are regarded as a topological solitons \cite{vortex1,vortex2,DS}. In other words, the dark soliton has a topological phase due to the phase difference induced by the wave function with asymmetry distribution \cite{DS}. Indeed, the topological field and classification of dark solitons have been clearly depicted \cite{Zhao1}. On the other hand, another common 1-D nonlinear excitation is the so-called bright soliton on the zero background \cite{DS,BS1,BS2}. However, finding the topological phase structure of bright solitons remains a challenge. The difficulty comes from that the bright solitons are generally considered to have the symmetry wave function and the constant phase distribution.

In this letter, we report the topological properties of bright solitons induced by the phase jump under the self-steepening. Importantly, this phase jump are affected by the arbitrary velocity of bright solitons. This is different from the well-known dark solitons \cite{DS,well-known1,well-known2}, here the phase jump is gradually increases to less than $\pi$ with the velocity of bright solitons increasing. In addition, we clearly depict the topological field of such bright solitons through the topological vector potential, for which consists of a series of point-like magnetic fields arranged in the single line. We find that these point-like magnetic fields corresponds to the density maximum of bright solitons and each elementary magnetic flux is $\pi$, in contrast to the 2-D topological excitation of a vortex that admits a zero-density core \cite{vortex2}. Surprisingly, we show that three lines of point-like magnetic fields are composed of two bright solitons, namely, an additional topological field can be generated between two bright solitons, except for the topological field possessed by each bright soliton. These results provided the possibility to study the topological phase of bright solitons in experiment.

We consider the \emph{generalized Chen-Lee-Liu} (generalized CLL) equation, which provides perhaps the most basic model system supporting topological excitation of bright solitons
\begin{eqnarray}
iu_t+\frac{1}{2}u_{xx}+\sigma|u|^2u+i\gamma|u|^2u_x=0,\label{gcll}
\end{eqnarray}
where $u(x,t)$ is the wave envelope, $x$ is the transverse variable and $t$ is the propagation variable. Coefficients $\sigma$ and $\gamma$ are arbitrary real constants is presented, which defines the term of the self-phase modulation (SPM) and self-steepening effect respectively. The feature of self-steepening without SPM can be described by the CLL equation when $\sigma=0$, and which can be observed through a Gaussian optical pulses of 110 fs passing a 8-mm-long $\beta$-barium metaborate with field intensities of a few $\rm{TW/cm^2}$ \cite{Moses}. Indeed, the combination of these two nonlinear terms early account for the asymmetric spectral of ultrashort pulses \cite{SS1,SS2,SS3}. Recently, the generalized CLL equation (\ref{gcll}) is used in the investigation of modulated wave dynamics of waves propagating through a single nonlinear transmission network \cite{network}, and provided the classification of possible wave structures propagating in a normal dispersion fiber \cite{waveSt}.

The generalized CLL equation (\ref{gcll}) is integrable. Namely, it admits various exact localized-wave solutions \cite{well-known2,generalCLL,soliton,RW, brea}. Here we consider the bright solitons on the zero background, which have nontrivial phase distribution \cite{soliton}. By using the method in Ref.\cite{generalCLL}, the bright soliton solution can be written as $u(x,t)=\frac{4w^2}{4w^2e^{-\alpha}+(\sigma-v\gamma+iw\gamma)e^{\alpha}}\exp{(i\theta)}$. By rationalizing the denominator, the above formula can be simplified as
\begin{eqnarray}
u(x,t)=\left[p(x,t)-iw\gamma\right]q(x,t)e^{i\theta},\label{1soliton}
\end{eqnarray}
with
\begin{eqnarray}
p(x,t)&=&4w^2e^{-2\alpha}+\sigma-v\gamma,\nonumber\\
q(x,t)&=&\frac{4w^2e^{-\alpha}}{w^2\gamma^2+p^2(x,t)},\nonumber
\end{eqnarray}
where $\alpha=w(x-vt)+s$, $\theta=vx-\frac{1}{2}(v^2-w^2)t$. The parameters $w$ and $v$ corresponds to soliton's width and velocity, the real constant $s$ is introduced in order to the maximum peak is located at the center. In contrast to the standard nonlinear Schr\"{o}dinger (NLS) equation \cite{DS,BS1,BS2}, Eq.(\ref{1soliton}) has an additional imaginary part related to the self-steepening. Therefore, we can speculate that this imaginary part makes the bright soliton induced by the self-steepening have similar topological properties to the dark soliton of NLS equation \cite{Zhao1}.

It is well-known that one measurable quantity of physical importance in exploring solitons is the phase. Such bright soliton solution admit the nontrivial phase as function of transverse coordinate by
\begin{eqnarray}
\phi(x)=\arg[u(x,0)],
\label{fx}
\end{eqnarray}
here we ignored the trivial phase factor $e^{i\theta}$ of Eq.(\ref{1soliton}). The magnitude of phase jump of the bright soliton from the integral expression can be calculated as
\begin{eqnarray}
\Delta\phi&=&\int^{+\infty}_{-\infty}d\phi(x)\nonumber\\
&=&\int^{+\infty}_{-\infty}-\frac{\gamma}{2}|u(x,0)|^2dx,
\label{df}
\end{eqnarray}
which is equivalent to the phase shift of the dark solitons on a nonvanishing background in Ref.\cite{ps-DS}, but the background intensity of such bright solitons is zero. Obviously, the above expression naturally degenerated to bright solitons with constant phase under standard NLS equation when $\gamma\rightarrow0$. Importantly, the phase distribution trend must be judged by the gradient of phase, which means the sign of the phase gradient correspond to the rise or fall of phase jump. We find that the direction of the phase jump can be described through the sign of self-steepening $\gamma$, which can be controlled through the wave vector mismatch \cite{SS3}.

To reveal the topological properties of bright soliton, we introduce a function $F[z]$ and extend into the complex plane ($z=x+iy$), which is the analytic extension of the integrand function in the area theorem, i.e.,
\begin{eqnarray}
F[z]=-\frac{\gamma}{2}|u(z,0)|^2,
\label{fz}
\end{eqnarray}
here we can simplify $F[z]=\mu[x,y]+i\nu[x,y]$. The vector potential $\textbf{A}$ along the real $x$ coordinate, which is introduced by considering a circle integral in the complex plane, i.e., $\oint_CF[z]dz=\oint\textbf{A}\cdot d\textbf{r}$, with
\begin{eqnarray}
\textbf{A}=\mu[x,y]\textbf{e}_x-\nu[x,y]\textbf{e}_y.\label{A}
\end{eqnarray}

From the expression of the integrand function in the area theorem, we know that $F[z]$ might have some singularities on the complex plane (denoted by $z_n=x_n+iy_n$ with $n$ as an integer) that corresponds to the maximum points of the density amplitude where the flow velocities diverge. We can then calculate the vector potential by determining the singularity locations in the complex plane, i.e.,
\begin{eqnarray}
x_n=0,~y_n=\pm\frac{1}{w} \arctan\left[\frac{w\gamma\sqrt{\zeta}}{1+(v\gamma-\sigma)\sqrt{\zeta}}\right]+\frac{n\pi}{w}.&&\label{sing}\\
(n=0, \pm1, ...)&&\nonumber
\end{eqnarray}
where $\zeta=\frac{1}{(\sigma-v\gamma)^2+w^2\gamma^2}$. This implies that the corresponding magnetic field $\textbf{B}=\nabla\times\textbf{A}$ will be zero everywhere in the whole complex plane except for those singular points.

According to the Residue theorem, each singularities corresponds to a quantized magnetic flux of elementary $\oint_CF[z]dz=\pm\pi$, where $\pm$ represents the magnetic flux direction. Moreover, such $\pi$ magnetic flux corresponding to a monopole with a charge of $1/2$ \cite{1/2}. In other words, these monopole emerge in pairs periodically, and merely located on the imaginary axis. It is similar to compare with the Aharonov-Bohm effect \cite{ABef} that predicts a topological phase when an electron moves around a solenoid, which the magnetic flux can be described by a Wess-Zumino topological vector potential \cite{Wess-Z}.

\begin{figure}[htbp]
\centering
\includegraphics[height=75mm,width=83mm]{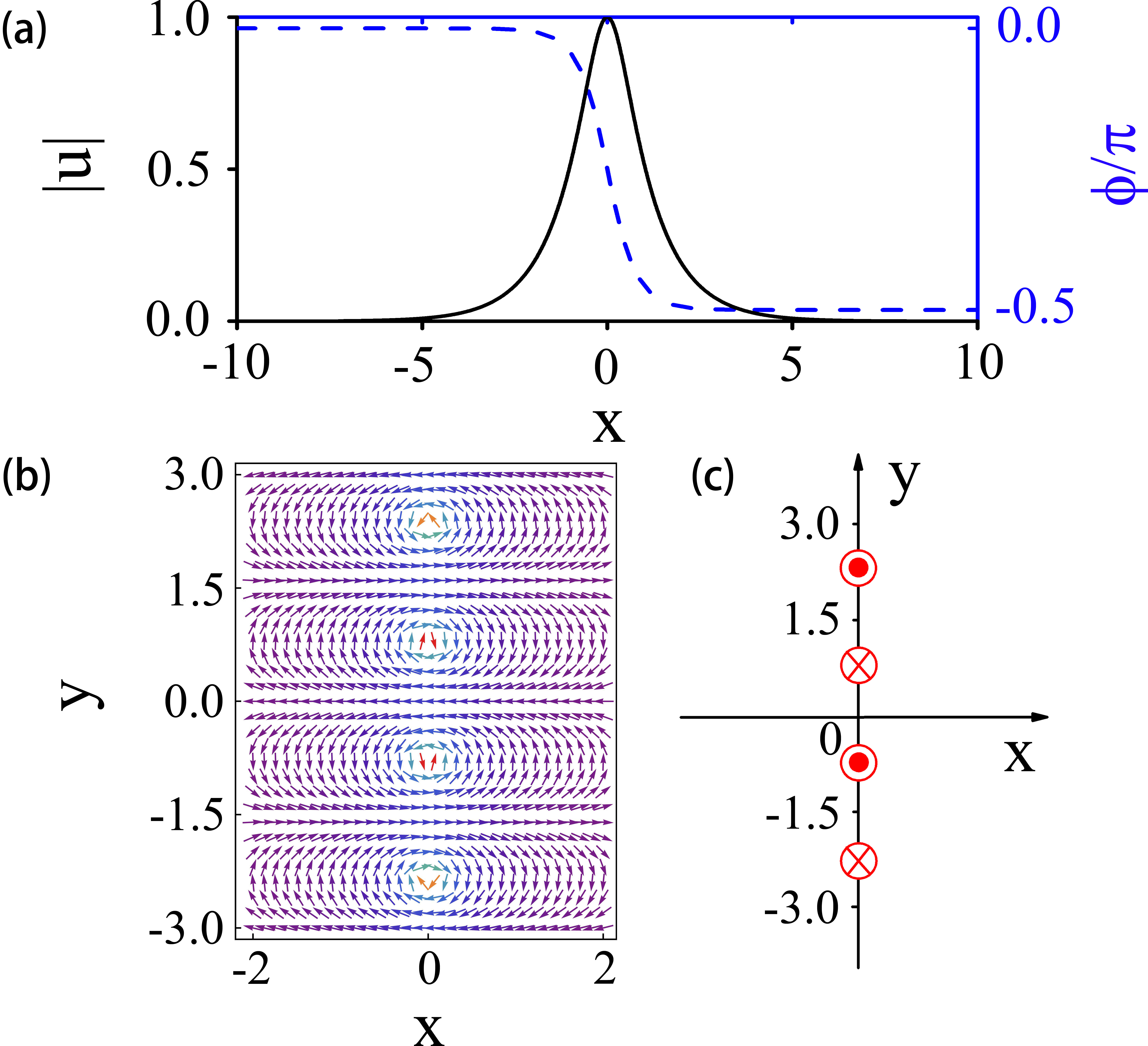}
\caption{Bright solitons with the topological phase, described by Eq.(\ref{gcll}). (a) Amplitude profile and phase distribution corresponds to the black solid line and the blue dotted line, respectively. (b) The topological vector potential $\textbf{A}$, and (c) the corresponding magnetic field $\textbf{B}$ for the bright soliton. Among them, the parameters of such bright soliton are $w=1,v=1/2, s=\ln2/2, \sigma=1$, and $\gamma=2$. The point-like magnetic fields are located at $(0,\pm0.79)$ with a period of $\pi$ along the y-axis for the bright soliton, and each elementary magnetic flux is $\pi$.} \label{fig1}
\end{figure}

Figure \ref{fig1} shows a particular case of bright solitons with topological phase properties when introduced the effect of self-steepening. We set the parameters of bright soliton solution in Eq.(\ref{1soliton}) as $w=1, v=1/2, s=\ln2/2$, and $\sigma=1, \gamma=2$, then the magnitude of phase jump $\Delta\phi=-\pi/2$ can be calculated by Eq.(\ref{df}). Such topological vector potential and effective magnetic field distribution corresponds to the maximum of the amplitude profile. Clearly, the 1-D bright soliton solution moving on the real axis cannot see the magnetic fields scattered on the complex plane, however, it will acquire a phase jump due to the presence of the vector potential. This effect results from the self-steepening similar to the topological properties of standard dark soliton \cite{Zhao1}. Moreover, the phase distribution is reversed then the vector potential and magnetic field are reversed accordingly when the signs of $v$ and $\gamma$ are changed at the same time.
\begin{figure}
\centering
\includegraphics[height=40mm,width=86mm]{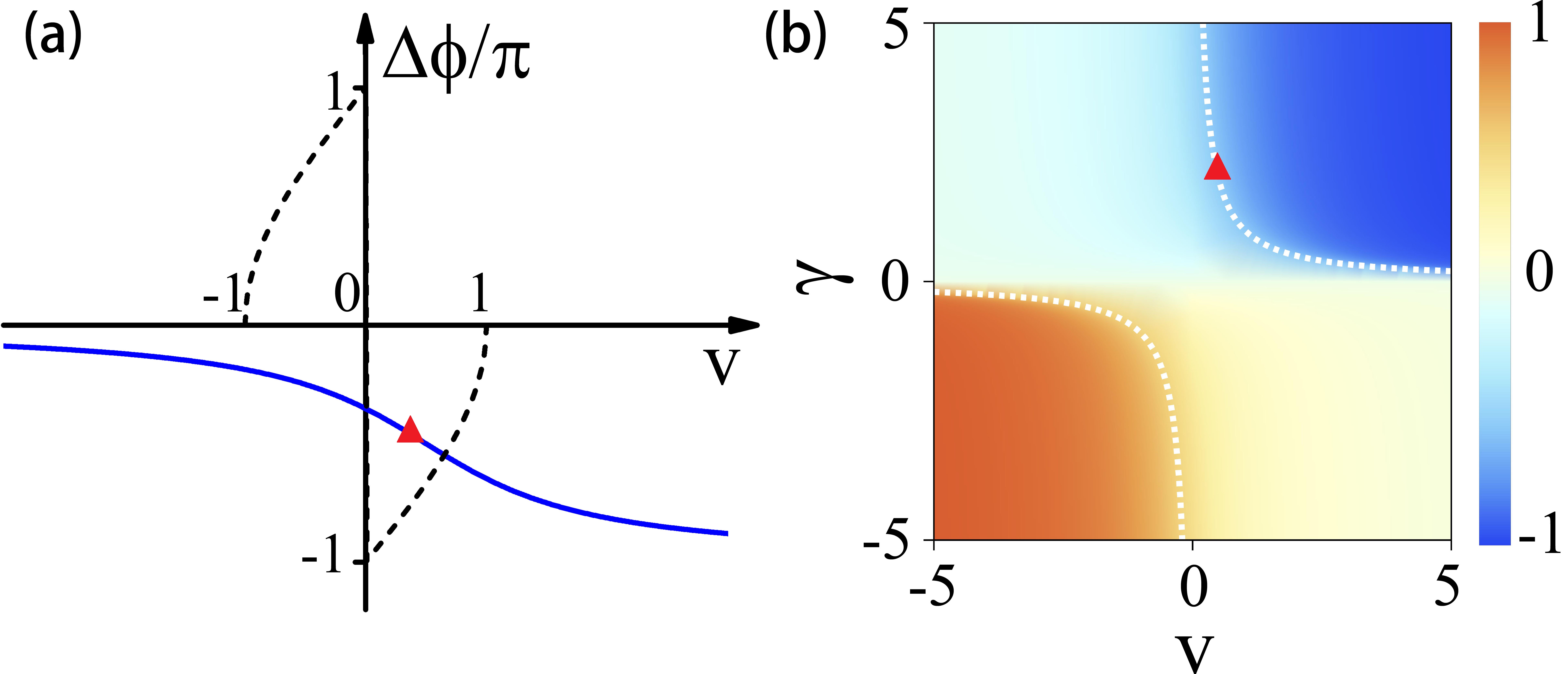}
\caption{(a) The effect of velocity $v$ on the phase jump $\Delta\phi$. (b) The influence of velocity $v$, self-steepening $\gamma$ on the phase jump $\Delta\phi$. It should be emphasized that the darker part of the color bar only tends to $\pm\pi$, moreover, $\Delta\phi=0$ corresponds to the general case of $\gamma=0$. The white dashed line corresponds to $|\Delta\phi|=\pi/2$, the red triangle corresponds to the parameters in Fig.\ref{fig1}, and the other parameters are the same as Fig.\ref{fig1}.} \label{fig2}
\end{figure}

It is shown that such bright soliton structures associated with the self-steepening effect have nontrivial phase jump. The velocity $v$ and self-steepening $\gamma$ plays a significant role in the magnitude of phase jump as can be seen from Fig.\ref{fig2} below. Figure \ref{fig2}(a) shows the case of phase jump depending on the bright soliton's  arbitrary velocity as shown by the blue solid line, where the phase jump is defined by Eq.(\ref{df}) and the sign of $\Delta\phi$ indicates the direction of phase jump. We choose the $\gamma=2$, that means the red triangle in this line graph corresponds to the case of $v=1/2$, $\Delta\phi=-\pi/2$ in Fig.\ref{fig1}. Obviously, $|\Delta\phi|$ increases gradually to approach $\pi$ as the velocity increases for the case of $v>0$, in contrast to the standard dark soliton \cite{Zhao1} (see the black dotted line). Meanwhile, we also find the self-steepening effect play an important role in the phase jump based on Eq.(\ref{df}). In other words, the phase jump is not only determined by the velocity but the self-steepening effect of such bright soliton. Therefore, the influence of two parameters $v$ and $\gamma$ on the phase jump is shown in Fig.\ref{fig2}(b), here the white dashed line represent $|\Delta\phi|=\pi/2$. Clearly, the sign of $\gamma$ can directly affect the directions of phase jump, where the orange and blue regions corresponds to the rise or fall of the phase distribution respectively. It is worth noting that $\Delta\phi=0$ for the case of $\gamma=0$ without the self-steepening effect, which means such well-known bright soliton has no phase jump then no topological properties.
\begin{figure}
\centering
\includegraphics[height=62mm,width=70mm]{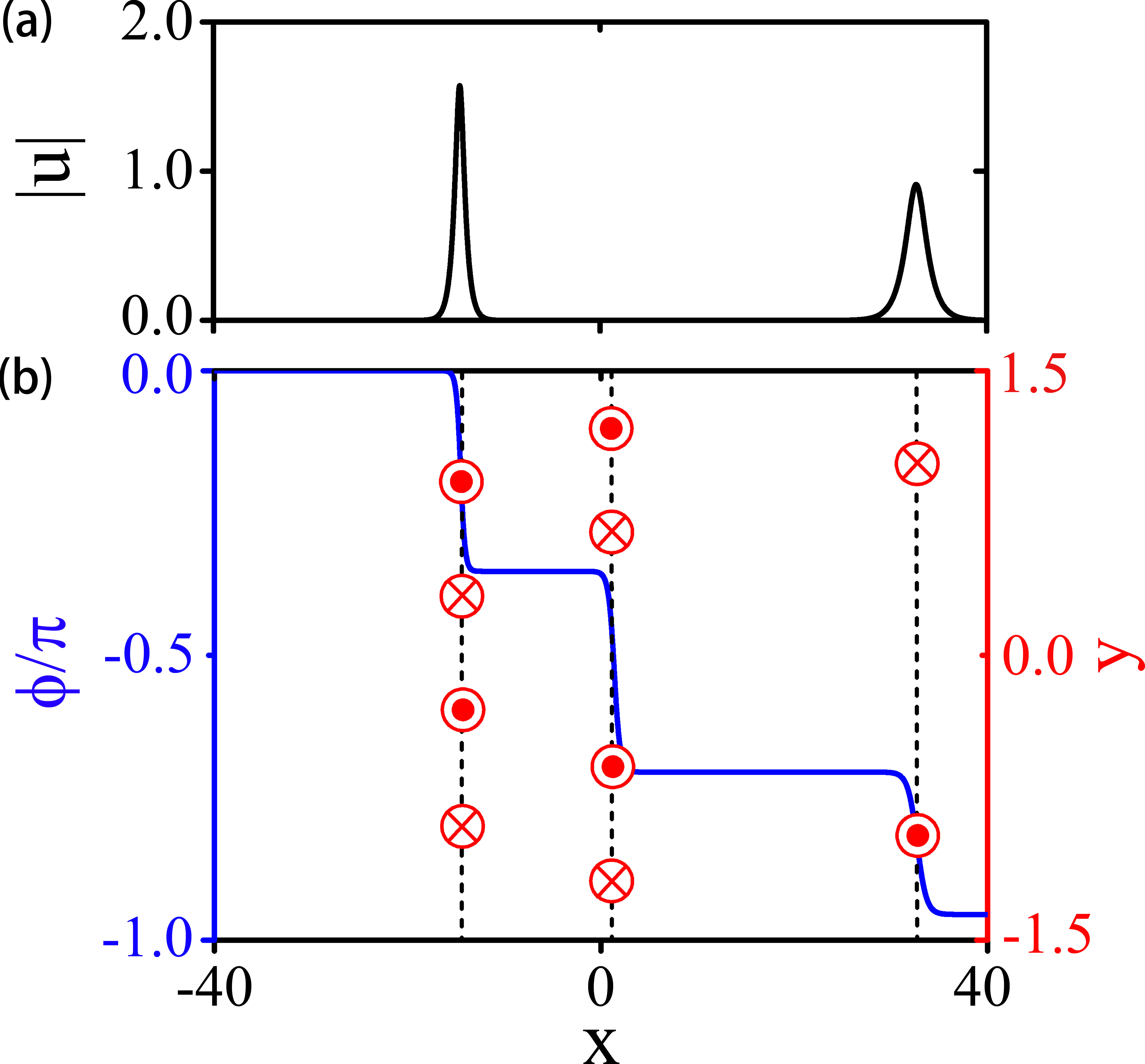}
\caption{The amplitude profile and the topological phase distribution of two-soliton with asymmetric structure at large separation. The magnetic field $\textbf{B}$ zero points are located at $(-14.5,\pm0.50)$ and $(32.7,\pm1.18)$ with a period of $\pi/2$ and $\pi$ along the y-axis, the phase jump of individual solitons are $\Delta\phi_1=-0.35\pi$ and $\Delta\phi_2=-0.25\pi$. Among them, the parameters are $v_1=v_2=0$, $w_1=1, w_2=2$, $\mu_1=-30, \mu_2=30$, $\theta_1=\theta_2=0$ and $\sigma=1, \gamma=1$. } \label{fig3}
\end{figure}

Similarly, we can also analyze the two bright solitons through the topological vector potential and the point-like magnetic field. The two-soliton solution of the generalized CLL equation (\ref{gcll}) can be presented through nonlinear superposition principle given by bilinear transformation \cite{generalCLL} as
\begin{eqnarray}
u^{[2]}(x,t)=\frac{\mathcal{F}_1}{\mathcal{F}_2},\label{2-soliton}
\end{eqnarray}
where
\begin{small}
\begin{align}
&\mathcal{F}_1=\nonumber\\
&[(1+\frac{\sigma\eta}{4w_2^2}-i\frac{\gamma\eta}{4w_2})\cosh{X_2}-(1-\frac{\sigma\eta}{4w_2^2}+i\frac{\gamma\eta}{4w_2})\sinh{X_2}]e^{iY_1}\nonumber\\
&+[(1+\frac{\sigma\eta}{4w_1^2}-i\frac{\gamma\eta}{4w_1})\cosh{X_1}-(1-\frac{\sigma\eta}{4w_1^2}+i\frac{\gamma\eta}{4w_1})\sinh{X_1}]e^{iY_2},\nonumber\\
&\mathcal{F}_2=(1-\zeta_1+i\gamma\sigma\zeta_2)\cosh{A_1}+(\sigma\beta_1+i\gamma\beta_3)\cosh{A_2}\nonumber\\
&-(1+\zeta_1-i\gamma\sigma\zeta_2)\sinh{A_1}+(\sigma\beta_2+i\gamma\beta_4)\sinh{A_2}\nonumber\\
&+[\frac{2\sigma}{(w_1+w_2)^2}+i\frac{\gamma}{w_1+w_2}]\cos{A_3}+\frac{(w_2-w_1)\gamma}{(w_1+w_2)^2}\sin{A_3},\nonumber\\
&X_1=w_1x+\mu_1, X_2=w_2x+\mu_2, Y_1=\frac{1}{2}w_1^2t, Y_2=\frac{1}{2}w_2^2t,\nonumber\\
&A_1=X_1+X_2,~ A_2=X_1-X_2, ~A_3=Y_1-Y_2.\nonumber
\end{align}
\end{small}with $\beta_1=(w_2^2+w_1^2)/4\omega_{12}^2$, $\beta_2=(w_2^2-w_1^2)/4\omega_{12}^2$, $\beta_3=(w_2+w_1)/4\omega_{12}$, $\beta_4=(w_2-w_1)/4\omega_{12}$, $\zeta_1=(w_1w_2\gamma^2-\sigma^2)\eta^2/16\omega_{12}^2$, $\zeta_2=(w_1+w_2)\eta^2/16\omega_{12}^2$, and among them $\eta=(w_1-w_2)^2/(w_1+w_2)^2$, $\omega_{12}=\omega_1\omega_2$. Specially, we introduced the constants $\mu_j$ which denote the initial locations of two bright solitons with zero velocities $v_1=v_2=0$. Importantly, the interaction behaviors of two bright solitons are controlled by the relative distance $(\Delta\mu=\mu_1-\mu_2)$, i.e., nonlinear interaction disappears approximately when relative distance $\Delta\mu$ is large. Figure \ref{fig3} shows an example fulfilling the constraint ($v_j=0$ and $\Delta\mu\gg0$), leading to independent propagation of two bright solitons.

As can be noted from Fig.\ref{fig3}, the separate propagation (for a weaker nonlinear interaction) of two bright solitons results in a special topological properties with nontrivial phase jump behavior. We find that this state of two bright solitons has three phase steps, two of them are the individual phase shift of solitons
\begin{eqnarray}
\Delta\phi_j=\int^{+\infty}_{-\infty}-\frac{\gamma}{2}\left[p_j^2(x,0)+w_j^2\gamma^2\right]q_j^2(x,0)dx,\nonumber
\end{eqnarray}
where $j=1,2$ represents the first and second bright solitons, $p_j(x,0)$ and $q_j(x,0)$ are given by Eq.(\ref{1soliton}).

Such phase jump structures can also described by topological fields (i.e., the topological vector potential and the effective magnetic fields). The triple-step structure of the phase jump and its corresponding magnetic fields on the complex plane for the two-soliton of Eq.(\ref{2-soliton}), as shown in Fig.\ref{fig3}(b). The distribution of point-like magnetic fields of two individual solitons are located at $\pm y_0$ with a period of $\pi/w$ along the y-axis, according to Eq.(\ref{sing}). Remarkably, an additional topological field are generated between two bright solitons, except for the topological field possessed by each bright soliton. In sharp contrast to the double jumps of the two dark solitons \cite{Zhao1}, the two bright solitons with self-steepening in Eq.(\ref{gcll}) can produce three phase jumps under nonlinear superposition.

In conclusion, bright solitons with self-steepening effect has an unexpected phase jump which can be explained by topological phase. Such topological phase can be described by the topological vector potential and effective magnetic field, which are similar to the dark solitons. We find that the velocity and self-steepening has great influence on the phase jump, in addition, the magnetic field and the direction of phase jump can be controlled through the sign of self-steepening. Moreover, multiple phase jumps can be generated by the nonlinear superposition of fundamental bright solitons, i.e., the three lines of the singular magnetic fields are composed of two bright solitons. Our research is helpful to broaden the understanding of bright soliton and provide a possibility for investigating the topological properties.

This work is supported by National Nature Science Foundation of China (Nos. 12022513, 11705145, 11947301, 11434013, and 11425522), the Major Basic Research Program of Natural Science of Shaanxi Province (Nos. 2017KCT-12 and 2017ZDJC-32), Natural Science Basic Research Plan in Shaanxi Province of China (Grant No. 2018JQ1003).
\end{CJK}
\end{document}